\documentclass[12pt]{article}

\usepackage{scicite}

\usepackage{times}
\usepackage{amssymb}
\usepackage{color}
\usepackage{amsmath}
\usepackage{amsbsy}
\usepackage{amsthm}
\usepackage{bbm}
\usepackage{bm}
\usepackage{epsfig}
\usepackage{lscape}
\usepackage{float}
\usepackage{subfigure}
\usepackage{hyperref}
\usepackage{ulem}
\usepackage{graphicx}

\newenvironment{sciabstract}{%
\begin{quote} \bf}
{\end{quote}}

\newcounter{lastnote}

\title{Levitated ferromagnetic magnetometer with energy resolution well below $\hbar$ }

\author
{Felix Ahrens$^{1,2}$, Wei Ji$^{3,4}$, Dmitry Budker$^{3,4,5}$, Chris Timberlake$^{6}$,\\
Hendrik Ulbricht$^{6}$, and Andrea Vinante$^{1,2 \ast}$\\
\\
\normalsize{$^{1}$CNR - Istituto di Fotonica e Nanotecnologie, 38123 Povo, Trento, Italy }\\
\normalsize{$^{2}$Fondazione Bruno Kessler (FBK), 38123 Povo, Trento, Italy}\\
\normalsize{$^{3}$Johannes Gutenberg University Mainz, 55128 Mainz, Germany}\\
\normalsize{$^{4}$Helmholtz-Institute, GSI Helmholtzzentrum für Schwerionenforschung, 55128 Mainz, Germany}\\
\normalsize{$^{5}$Department of Physics, University of California at Berkeley, Berkeley, California 94720-7300, USA}\\
\normalsize{$^{6}$School of Physics and Astronomy, University of Southampton, UK}\\
\\
\normalsize{$^\ast$To whom correspondence should be addressed; E-mail:  anvinante@fbk.eu}
}

\date{}

\begin{document} 


\baselineskip24pt


\maketitle

\begin{sciabstract}
   A quantum limit on the measurement of magnetic field has been recently pointed out, stating that the so-called Energy Resolution $E_\mathrm{R}$ is bounded to $E_\mathrm{R} \gtrsim \hbar$. This limit holds indeed true for the vast majority of existing quantum magnetometers, including SQUIDs, solid state spins and optically pumped atomic magnetometers. However, it can be surpassed by highly correlated spin systems, as recently demonstrated with a single-domain spinor Bose-Einstein Condensate. Here we show that similar and potentially much better resolution can be achieved with a hard ferromagnet levitated above a superconductor at cryogenic temperature. We demonstrate $E_\mathrm{R}=\left( 0.064 \pm 0.010 \right) \, \hbar$ and anticipate that $E_\mathrm{R}<10^{-3} \, \hbar$ is within reach with near-future improvements. This finding opens the way to new applications in condensed matter, biophysics and fundamental science. In particular, we propose an experiment to search for axionlike dark matter and project a sensitivity orders of magnitude better than in previous searches.
\end{sciabstract}


\section*{Introduction}

Magnetic fields can be measured through a variety of techniques, each one with different advantages and drawbacks. Some specific applications require pushing the magnetic field resolution towards extreme values: this is the case of weak biomagnetic fields from human brain, of the order of several fT \cite{MEG2012,MEG2017} or fundamental physics, where even lower pseudomagnetic fields down to $10^{-20}-10^{-25}$\,T can be expected from axionlike dark matter \cite{Fadeev2021,Vinante2021} or relativistic frame-dragging on quantum spins\cite{Fadeev2020gravity}. This raises the question: what is the smallest magnetic field that can be detected for a given temporal and spatial resolution? 

A recent review \cite{Mitchell2020colloquium} discussed a figure of merit to compare different magnetic field sensors, the Energy Resolution per unit bandwidth \cite{Robbes2006}:
\begin{equation}
  E_\mathrm{R}=\frac{S_{B}V}{2 \mu_0}\,,
\end{equation}
where $S_{B}$ is the power spectral density (PSD) of the magnetic field noise, $V$ is the sensor ``averaging'' volume and $\mu_0$ is the vacuum permeability. $E_\mathrm{R}$ embodies an effective combination of magnetic field, spatial and temporal resolution and has the dimension of action. By setting $E_\mathrm{R} \geq\hbar$ one formally defines a quantum limit, which has been dubbed in literature Energy Resolution Limit (ERL) \cite{Mitchell2020colloquium}. Several classes of magnetometers are indeed subject to standard quantum limits, which can be expressed as $E_\mathrm{R}>\alpha \hbar$ with $\alpha$ being of the order of unity, therefore reducing to the ERL. In particular, under fair assumptions all most sensitive practical magnetometers are subject to the ERL, including dc superconducting quantum interference devices (SQUIDs) \cite{Tesche1977,Koch1980,Awschalom1988,Falferi2008}, optically pumped atomic magnetometers \cite{Romalis2003, Jimenez2017} and arrays of independent nitrogen-vacancy (NV) spins in diamond \cite{Zhou2020,Mitchell2020}.

However, there is no general derivation of the ERL from first principles \cite{Mitchell2020colloquium}. Standard quantum limits on measurement precision typically arise from the uncertainty principle \cite{Braginsky1975,Caves1982}, applied to the physical observable that is monitored to infer the magnetic field. Instead, in the definition of ERL there is no reference to any specific physical observable.

Indeed, several solutions have been envisioned as potentially able to significantly surpass the ERL. Notably, $E_\mathrm{R}=0.075 \, \hbar$ has been recently achieved with a single-domain spinor Bose-Einstein Condensate thus overcoming the ERL by more than a factor 10 \cite{Mitchell2022}, while a single NV center manipulated with multiple quantum techniques is claimed to have achieved $E_\mathrm{R}=0.18 \, \hbar$ \cite{Zhao2022}. This confirms that the ERL is not fundamental, and suggests that a new class of ultrasensitive magnetometers can be developed.


Here, we demonstrate resolution well below the ERL, $E_\mathrm{R}= \left( 0.064 \pm 0.010 \right) \, \hbar$ by levitating a micron-scale ferromagnet in a superconducting trap, and inferring the applied magnetic field through the mechanical torque induced in the ferromagnet. Remarkably, this concept of magnetometer is closely related to the first compass-based magnetometer ever reported in history, developed by Carl Friedrich Gauss in 1833 \cite{Gauss1833}, consisting of a torsion pendulum with a ferromagnetic needle as inertial member. The essential feature that allows our remarkably simple system to overcome the ERL is the nature of a hard ferromagnet: a macroscopic number of spins locked in the same direction by the exchange interaction and therefore forced to rotate altogether. At the slow rotational frequencies involved, the internal spin dynamics is effectively negligible, and quantum noise is rapidly averaged out \cite{Kimball2016}. 
Our result opens new prospects in research areas requiring ultrahigh sensitivity and high spatial resolution magnetometry, such as the detection of biomagnetic fields from single events in the brain \cite{MacGregor2012} or in fundamental physics. In particular, our setup is highly insensitive to external magnetic noise due to the shielding from the superconducting trap, while maintaining high sensitivity to different spin-dependent fields arising from new physics beyond the standard model, such as exotic spin-dependent forces \cite{Fadeev2021,Vinante2021} and axionlike dark matter\cite{Kimball2023}, or relativistic gravitomagnetic effects \cite{Fadeev2020gravity}.
As a case study, we estimate the potential reach of this technique in the direct search for axion dark matter with coupling to electron spins. 

\section*{Experimental setup}

The setup is sketched in Fig.\,1A. The magnetic field sensor is a hard ferromagnetic microsphere picked from a commercial NdFeB alloy powder \cite{Magnequench}. This micromagnet is placed on the bottom of a superconducting trap machined from a bulk piece of lead with a hemispherical bottom section. As the lead trap is cooled below the superconducting transition temperature, the repulsion of the magnet from the superconductor due to the Meissner effect lifts the micromagnet and, in combination with gravity, traps it in a stable position. At equilibrium the magnetic dipole of the micromagnet lies on the horizontal plane. The translational and rotational motion of the magnet are detected with a dc SQUID sensor \cite{QuantumDesign} via a superconducting 8-shaped pick-up coil. The whole setup is enclosed in a vacuum chamber with a moderate magnetic shielding (factor of 10) and cooled in liquid helium to temperature of $T=4.2$ K. 
\begin{figure}[!t]
\centering
\includegraphics[width=0.6 \linewidth]{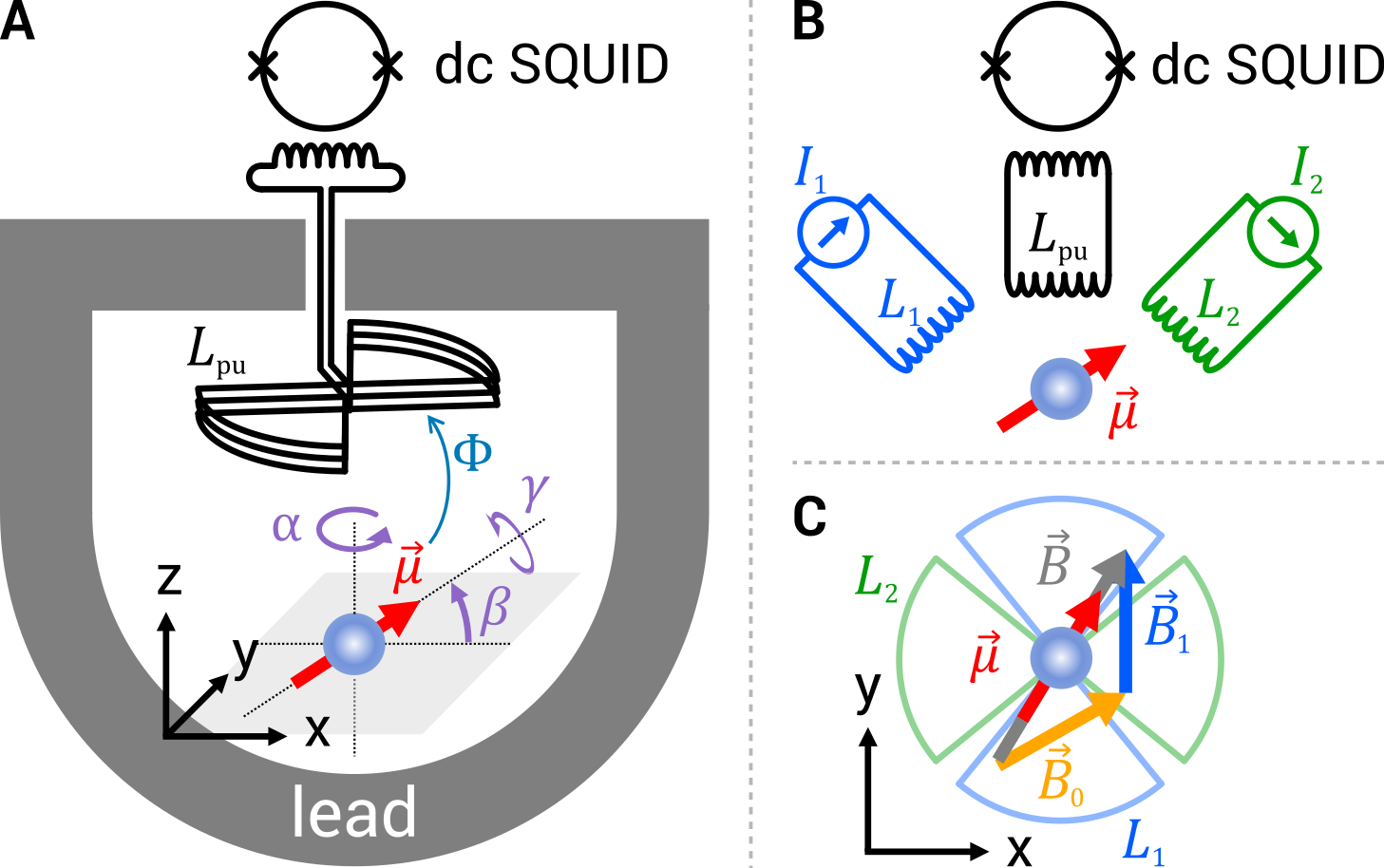}
\caption{(A) Experimental setup. A spherical hard micromagnet is levitated inside a lead (type I superconducting) trap with hemispherical bottom. The motion of the micromagnet in its different degrees of freedom is measured with a dc SQUID sensor through the magnetic flux $\Phi$ coupled into a gradiometric pick-up coil $L_\mathrm{pu}$. For the sake of clarity, the actuator coils $L_1$ and $L_2$, which serve for the generation of bias magnetic fields, are omitted. (B) Circuit diagram of the experimental setup consisting of one pick-up coil $L_\mathrm{pu}$ and two actuator coils $L_1$ and $L_2$ which can be used for applying bias magnetic fields. All coils are 8-shaped to maximize coupling to horizontal rotations and placed approximately $2$ mm above the magnet. (C) The effective magnetic field $\mathbf{B}$ on the horizontal plane, which determines the resonance frequency $\omega_\alpha=\sqrt {\mu B/I}$ of the $\alpha$ mode, is the vector sum of a constant field $\mathbf{B_0}$ frozen in the superconducting trap, and an additional field $\mathbf{B_1}$ generated by the coil $L_1$ placed above the magnet. 
}
\end{figure}

We can identify the normal modes of the micromagnet motion through comparison with analytical and numerical modelling of the system \cite{Vinante2020,Timberlake2019,Lin2006,Vinante2022}. Furthermore, combined measurements of the mode frequencies allow for an in-situ estimation of the radius $R=\left(20.78  \pm 0.20 \right) \, \mu$m and magnetization $\mu=\left( 6.91 \pm 0.18 \right) \times 10^{5}$ A/m of the micromagnet \cite{Supplementary}. This is important, since the parameters, in particular the magnetization, may change when cooling to low temperature. 

Here, we focus on the so-called $\alpha$ mode, which corresponds to a rotation in the horizontal plane. Because of cylindrical symmetry, one would naively expect the $\alpha$ mode to be a free rotation. Instead, we observe it as a trapped librational mode, as in a torsion pendulum, with resonance frequency of the order of $90-150$\,Hz, which fluctuates upon repetition of the cooldown procedure, 
suggesting that a residual static magnetic field is left in the trap. This could be due either to magnetic flux trapped in the pick-up coil circuit, or to penetration of external field due to mixed-state effects in the trap. 
The horizontal component $\mathbf{B_0}$ locks the magnet in a fixed direction, yielding for the $\alpha$ mode a rotational stiffness $\mu B_0$ and an angular resonant frequency $\omega_\alpha=\sqrt {\mu B_0/I}$ where $I$ is the moment of inertia. 
\begin{figure}[!t]
\centering
\includegraphics[width=0.5\linewidth]{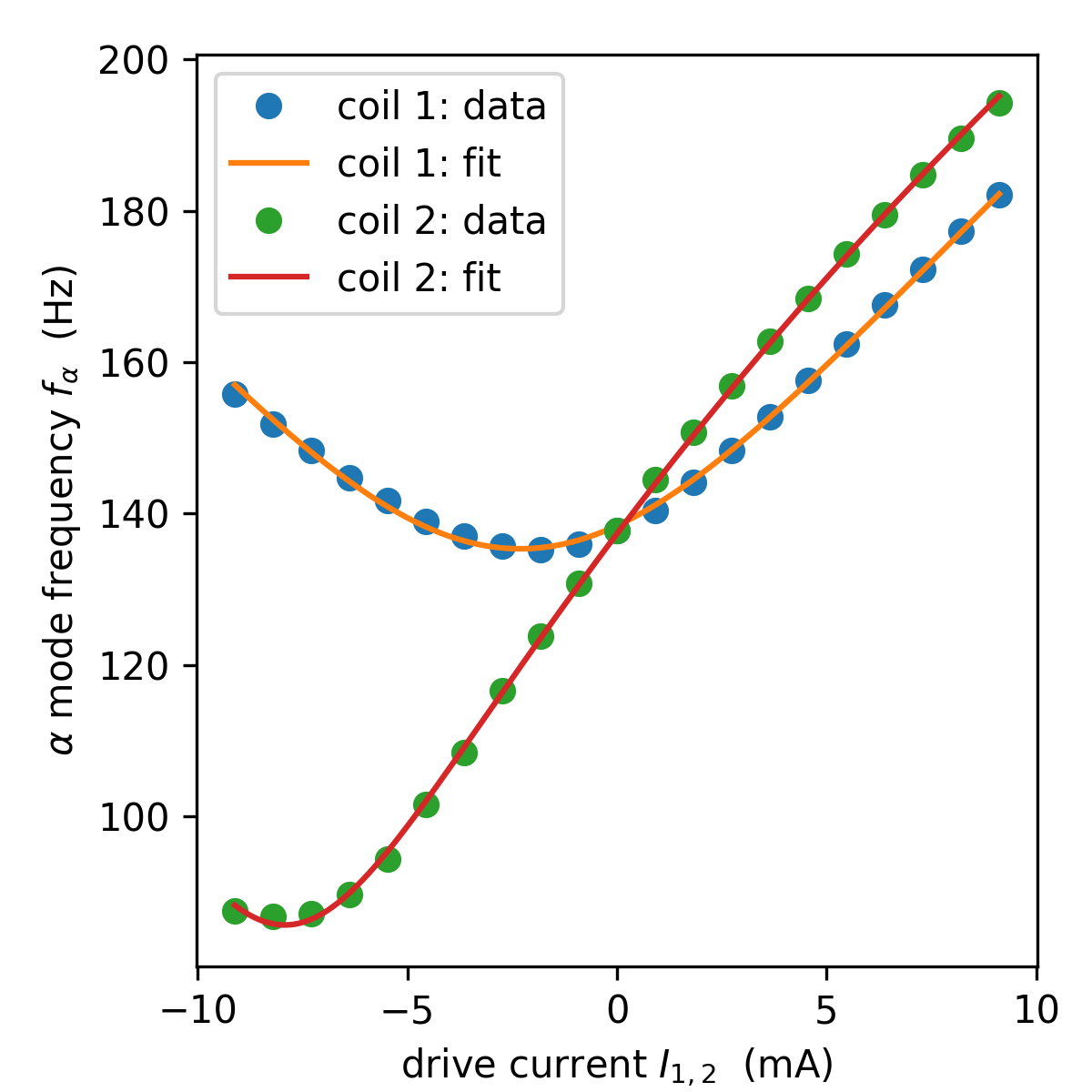}
\caption{Frequency of the $\alpha$ mode $f_\alpha=\omega_\alpha/2\pi$ as a function of the current in the field coils $1$ and $2$. The data are fitted with a model assuming a static residual field in the superconducting trap with horizontal component $B_0$. By design, the fields from the two coils are expected to be horizontal at the micromagnet position and orthogonal to each other. 
}
\end{figure}
We can apply additional horizontal magnetic fields $\mathbf{B_1}$ and $\mathbf{B_2}$ in two orthogonal directions, say $x$ and $y$, via a pair of 8-shaped coils placed approximately $2$ mm above the equilibrium position.
When we apply the field $\mathbf{B_{1}}$ (or similarly $\mathbf{B_{2}}$), the total static field changes in amplitude and direction, leading to a shift in the $\alpha$ mode resonance frequency to $\omega_\alpha=\sqrt {\mu B/I}$ where $B=|\mathbf{B_0}+\mathbf{B_1}|$. Figure 2 shows the frequency of the $\alpha$ mode as a function of the current applied to each coil, together with a fit with the theoretical model assuming a residual static field $B_0$. The data are in good agreement with the simple model. From the fit we infer $B_0$, the angle $\theta_i$ between the field applied from each coil and the coil coupling factors $\lambda_i=B_i/I_i$ where $I_i$ is the current sent through each coil. 

\section*{Magnetic field resolution}
In the following we shall consider only coil 1, which generates a field nearly orthogonal to $\mathbf{B_0}$. 
The values from the fit in Fig. 2 are ${B_0=\left( 1.38 \pm 0.21 \right)}$\,$\mu$T, $\theta_1=\left(73.0 \pm 0.2 \right)$ deg and ${\lambda_1=\left( 180 \pm 28 \right)}$\,$\mu$T/A. 

Instead of applying a dc field we consider now a sinusoidal field $\mathbf{B_1} \left( \omega \right)$ with frequency $\omega$ around the resonance frequency $\omega_0$ of the $\alpha$ mode. This generates a torque $\tau=\mu B_1 \mathrm{sin} \theta_1 $ which drives the $\alpha$ mode and is detected as an oscillation of the magnet. By measuring the smallest detectable $B_1$ field in a given frequency bandwidth or integration time, we can estimate the Power Spectral Density (PSD) of the magnetic field noise $S_B$, and thus the energy resolution $E_\mathrm{R}$ referred to the micromagnet volume $V$ through Eq.\,(1).

Theoretically, we can estimate the expected $S_B$ following the modelling of Ref.\,\cite{Vinante2021}. Assuming that we are in the thermal noise limit, the torque PSD is given by $S_{\tau}=4 k_\mathrm{B} T I \omega_0/Q$, where $Q$ is the mechanical quality factor.
The magnetic field power spectral density is then calculated as: $S_B=S_{\tau}/\left( \mu \, \mathrm{sin}\theta_1\right)^2$.
We perform our measurements at a pressure of $P=1.1 \times 10^{-3}$\,mbar, where we have separately estimated through a ringdown measurement the resonance frequency and quality factor of the $\alpha$ mode as $f_\alpha=\omega_\alpha/2 \pi=137.625$ Hz and $Q=3.96 \times 10^4$ \cite{Supplementary}. 

\begin{figure}[!t]
\centering
\includegraphics[width=\linewidth]{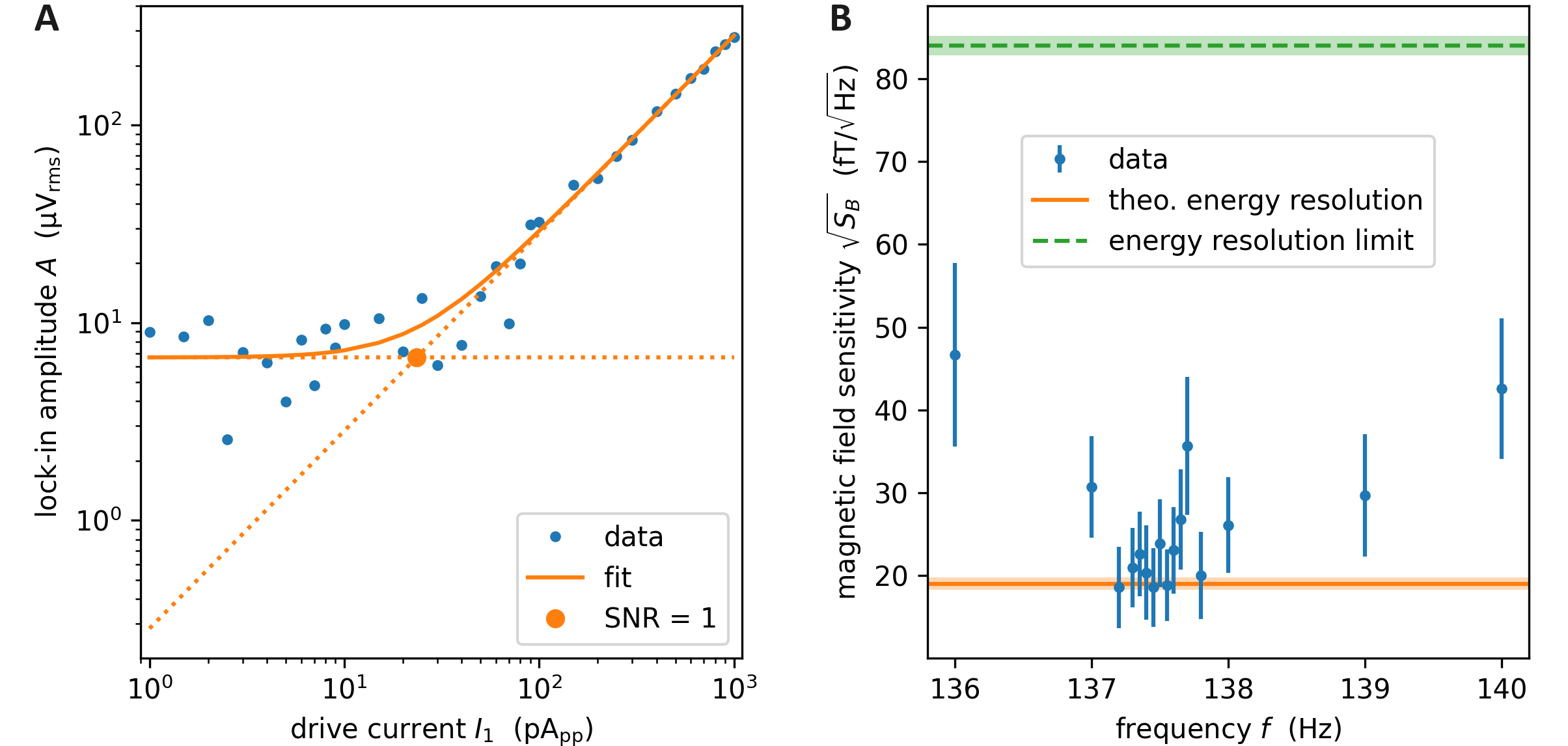}
\caption{(A) Typical magnetic field resolution measurement. The magnetic field $B_1$ is applied by sending a drive current $I_1$ through coil 1 at a frequency $\omega \approx \omega_\alpha$, the resonance frequency of the $\alpha$ mode. The quantity measured on the y-axis is the SQUID output signal measured synchronously with the driving by a narrowband lock-in amplifier. In the linear high current regime the torque induced by the field drives the $\alpha$ mode above thermal noise, whereas the signal drops below the noise at lower current. The magnetic field resolution is determined by the crossing point between the two regimes, which corresponds to SNR=1. (B) Magnetic field resolution, normalized as power spectral density per unit bandwidth $S_B$, at different frequencies around resonance. Close to resonance the data are in agreement with the thermal torque noise prediction (orange bottom solid line). The ERL referred to the volume of the magnet $V$ is also plotted (green dashed line). 
}
\end{figure}

Figure 3A shows a typical measurement of minimum detectable field as a function of the current applied to the coil at a fixed driving frequency $f$. Here we measure the mechanical response of the $\alpha$ mode with a lock-in amplifier synchronous with the driving, with noise bandwidth $\Delta f = 1/64$\,Hz. The plot clearly shows the transition from a linear regime, where we observe the driven oscillation, to a noise regime where the driven signal drops below the magnetic field noise $B_\mathrm{n}$. The intersection between the two regimes defines the current, and therefore the applied field, at which the signal to noise ratio (SNR) is equal to 1, i.e. $B_\mathrm{n}=\lambda_1 I_{1,\mathrm{SNR}=1}$. We can finally express the minimum field as power spectral density $S_B$ by taking into account that the noise power integrated over the measurement bandwidth is $B_\mathrm{n}^2=S_B \Delta f$.

The measurement is repeated at different frequencies around $f_\alpha$. The corresponding values of $S_B$ are shown in Fig. 3B. On the same figure we plot both the prediction of $S_B$ assuming the system is thermal noise limited and the ERL from Eq.\,(1). Details on the estimation of all experimental parameters and uncertainties are reported in the Supplementary Material. We note that the experimental values of $S_B$ are consistent with thermal noise only close to resonance. When the driving frequency is moved off-resonance the mechanical response is decreased so that the SQUID detection noise becomes significant.

\section*{Discussion and outlook}

Figure 3B shows that the magnetic field resolution averaged around resonance is in good agreement with the value expected from the thermal torque noise, and significantly better, a factor of 4.0 in field, with respect to the ERL referred to the micromagnet volume $V$. If we convert the minimum detected value of $S_B$, corresponding to the thermal noise limit, into an energy resolution, we find $E_\mathrm{R}=\left( 0.064 \pm 0.010 \right) \, \hbar$. This figure is comparable and slightly better than the one estimated for the spinor BEC reported in Ref. \cite{Mitchell2022}. This makes our magnetometer the most sensitive ever reported in literature in terms of bare energy resolution per unit volume.

One may object that the value of $E_\mathrm{R}$ is strongly dependent on the choice of the reference volume $V$ in Eq.\,(1), which is to some extent arbitrary. Should we take the micromagnet volume, or rather a larger volume including the field coil or the trap volume? Since we measure the magnetic field through the mechanical response of the magnet, which is determined by the actual field felt by spins in the magnet, we argue that only the magnet volume $V$ is relevant. This is indeed the same concept of energy resolution used in recent literature \cite{Mitchell2020,Mitchell2022}. It does not make any assumption on the spatial size of the field source, and therefore represents an intrinsic property of the sensor.


Another possible objection is that the setup involves a shielded trap, so it cannot be used as a practical magnetometer to measure field sources outside the trap. However, this issue can be circumvented by transferring the field to be measured from outside to inside the trap via a superconducting pick-up coil, as typically done with SQUID magnetometers. In this case, the geometry and the coupling of the pick-up coil would need to be optimized and this would call for a different figure of merit referred to the pick-up coil which could be defined as coupled energy resolution, in analogy with SQUID magnetometers  \cite{Clarke1979}.

The main actual limitation of the current setup is that high sensitivity is achieved only within a limited bandwidth around the resonance frequency. Indeed, one can think of our setup as a SQUID magnetometer where the levitated micromagnet acts as a narrowband resonant preamplifier, converting the input B field into a much higher field around $f_\alpha$. 
In order to widen the useful bandwidth, currently a few Hz around $f_\alpha=137.625$\,Hz, one can either increase the coupling to the SQUID or reduce the SQUID noise. 
A possible strategy in this direction is to implement a chip trap, which would enable a reliable miniaturization. We plan to do this in the near future. Eventually, the SQUID itself will become a fundamental limitation to the energy resolution since classical or quantum back-action noise from the SQUID will eventually dominate over the thermal noise upon sufficient increase of the coupling. In this case, the ultimate detectable torque is set by the Standard Quantum Limit on mechanical measurements \cite{Braginsky1975,Giffard1976,Caves1982,Vinante2021}. 

A scenario with an optimized quantum limited SQUID sensor was analyzed in Ref.\,\cite{Vinante2021}. It was shown that wideband operation can be achieved with a resolution in magnetic field four orders of magnitude below the ERL. For the current setup we can readily estimate the SQUID back-action noise. Quantum back-action would correspond to a PSD of the current induced in the pick-up coil $S_\mathrm{i}\approx 2 \hbar L_\mathrm{i}/L_\mathrm{t}^2$ \cite{Tesche1977, Auriga2006, Falferi2008} where $L_\mathrm{i}=1.8$\,$\mu$H is the SQUID input inductance and $L_\mathrm{t} \approx L_\mathrm{i}$ is the total inductance of the pick-up loop circuit. Assuming that the pick-up coil has the same coupling of the actuation coils, this translates into a magnetic field PSD of $S_B^{1/2}=1.7$\,aT/$\sqrt{\mathrm{Hz}}$ corresponding to $E_\mathrm{R}=4 \times 10^{-10} \, \hbar$. Even assuming a back-action noise $10^3$ times larger, more realistic for the commercial SQUID currently implemented \cite{Auriga2006}, the back-action noise would be $50$ aT/$\sqrt{\mathrm{Hz}}$. This indicates that, if narrowband operation is acceptable, a potentially enormous reserve for improvement in energy resolution exists. In the near future, operation at lower pressure where $Q>10^7$ has been already demonstrated \cite{Vinante2020} should allow improvement by more than a factor $10$ in magnetic field, corresponding to $E_\mathrm{R} < 10^{-3} \, \hbar$. Further improvements can be obtained by cooling to millikelvin temperatures.

\begin{figure}[!t]
\centering
\includegraphics[width=0.6\linewidth]{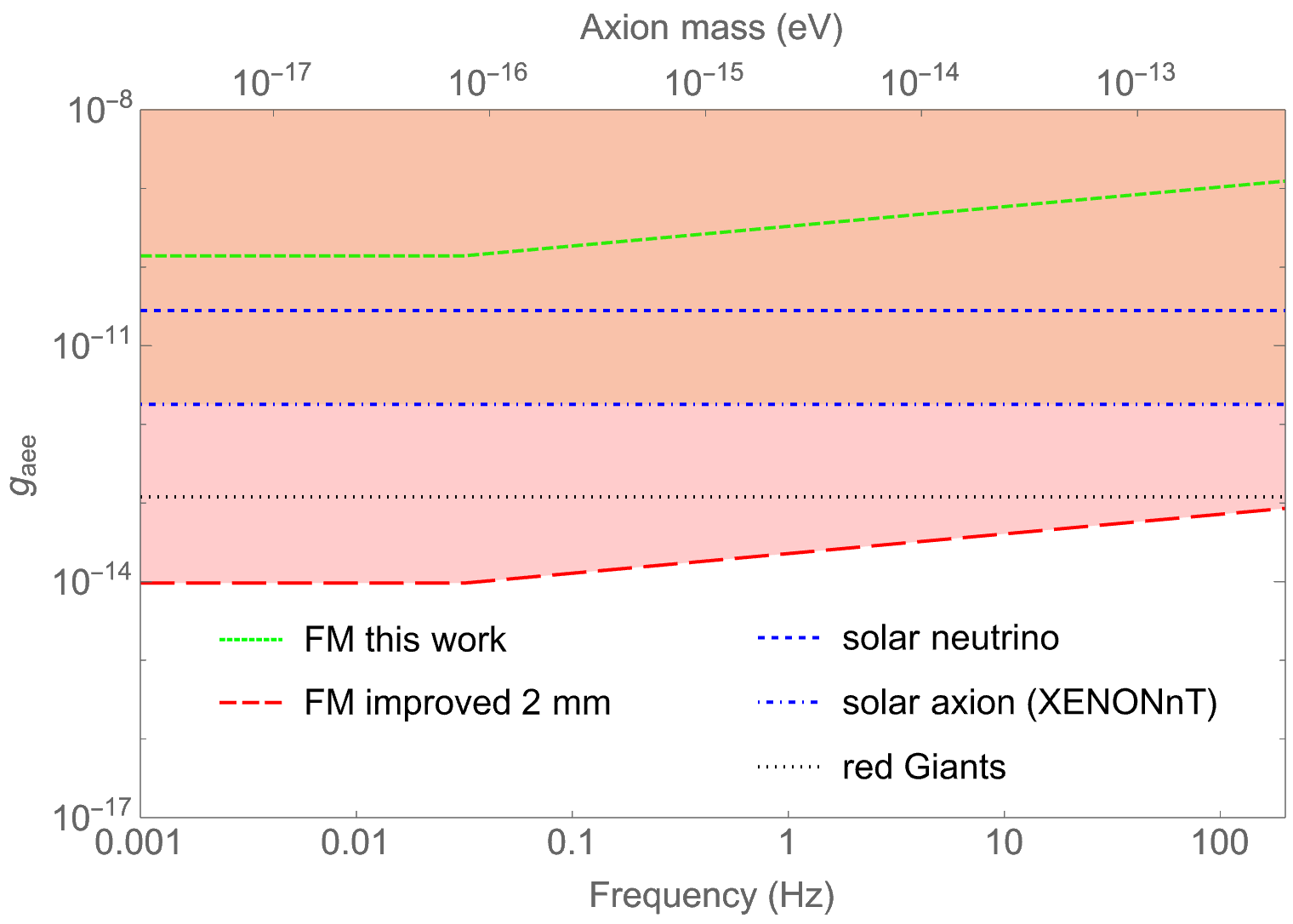}
\caption{Expected sensitivity for the axion-electron coupling constant $g_{aee}$ as a function of frequency. Green-dashed and red-dashed lines refer to the Ferromagnetic Magnetometer\,(FM) experiment of this work and to an improved larger FM ($R=2$ mm, $T=50$ mK, $Q=10^8$) for one year integration time at the thermal noise limit. 
The other three curves refer to existing astronomical limits on $g_{aee}$ from solar neutrinos \cite{gondolo2009solar}, solar axions\,\cite{aprile2022search} and from red Giants\,\cite{capozzi2020axion}. Note that our experiment will directly probe galactic-halo dark matter.}
\end{figure}

Finally, we note that our detector is narrowband but the frequency is tunable comfortably between mHz and kHz with bias fields or by modifying the trap shape. In addition, it is intrinsically insensitive to external magnetic field noise, owing to the superconducting trap shielding,  while maintaining high sensitivity to different spin-dependent fields which could arise from new physics beyond the standard model, such as axion dark matter fields \cite{Kimball2023} or exotic interactions arising from such fields \cite{Fadeev2021}. Here, we propose an experiment to search for axion dark matter predicted by the galactic halo model (i.e., a haloscope) exploiting the coupling of axions to the electron spins of the ferromagnetic magnetometer. 

Axions and axionlike particles (ALPs), here generically referred to as axions, are promising dark matter candidates that can have the correct relic abundance to form the main contribution of the galactic dark matter halo \cite{duffy2009axions,marsh2016axion}. Because axions are ultralight, they behave as an effective wavelike field and their gradient coupling to electron spins can be treated as an effective magnetic field acting on the ferromagnetic sensor with an amplitude in natural units \cite{Supplementary}: 

\begin{equation}\label{axionfield}
\mathbf{B}_{a}= \frac{g_{aee}}{2 e}\sqrt{2 \rho_a} \mathbf{v}_a\,,
\end{equation}

\noindent where $\rho_a\simeq0.4 \, \rm{GeV/cm}^3$ is the dark matter energy density in the Milky Way halo \cite{de2021dark}, $g_{aee}$ is the axion-electron coupling constant, $e$ is magnitude of the electron charge,
 and $\mathbf{v}_a\simeq10^{-3}c$ is the local galactic virial velocity. 
Note that 
according to Eq.\,(\ref{axionfield}), axion dark matter gradient interacts directly with electron spins as effective magnetic field, while in other proposals
superconducting objects act as transducers from axion to real magnetic field and spin-independent magnetometer can be used
\cite{higgins2023maglev}; our sensor is sensitive to both types of couplings.

In Fig.\,4, we estimate the potential reach of a search for cosmic dark matter on coupling constant $g_{aee}$ using a ferromagnetic magnetometer (FM) with parameters as in this work and with improved ones, for an integration time of one year. For simplicity, we consider thermal noise limiting the FM sensor performance, but we anticipate that SQUID readout noise can be kept lower than thermal noise in both configurations considered \cite{Vinante2021}. The improved configuration refers to a larger particle with radius $R=2$ mm cooled to $T=50$\,mK, with $Q=10^8$. The slight slope above 0.1 Hz is related with the axion coherence time. Further details on the derivation of the upper dark matter bounds are in the Supplementary Material \cite{Supplementary}. 
FM detectors are complementary to other haloscope experiments
which probe axion-electron coupling at a lower\,\cite{terrano2019constraints} or much higher mass range~\cite{crescini2020axion,fu2017limits}. Note that existing limits on axions (see website \cite{AxionLimits} for updated results) in our targeted mass range are from indirect astrophysical observation\,\cite{capozzi2020axion,gondolo2009solar,giannotti2017stellar} or from solar axions\,\cite{aprile2022search,fu2017limits}. Our proposed FM experiment will directly 
probe the galactic-halo dark matter with comparable or better sensitivity.

In conclusion, we have shown that a levitated-ferromagnet torque-based narrowband magnetic sensor surpasses the Energy Resolution Limit, with measured energy resolution of $E_\mathrm{R}=\left( 0.064\pm 0.010 \right)\hbar$ and $E_\mathrm{R}<10^{-3} \, \hbar$ being within reach with current parameters at lower gas pressure. This results opens the way to novel ultrasensitive measurements of magnetic fields or pseudomagnetic fields. In particular, this class of sensors is promising in the search for beyond-standard-model  exotic spin-spin interactions \cite{Fadeev2021,Vinante2021} and for direct detection of axionlike dark matter \cite{Kimball2023}. Future technological improvements, including on-chip miniaturization, optimized SQUIDs and operation at millikelvin temperatures, will enable further applications in fundamental physics as well as in other fields, such as biomagnetism \cite{MacGregor2012}.

\bibliography{references}
\bibliographystyle{Science}

\section*{Acknowledgments}

\paragraph*{Funding:}
We thank D. Grimsey for expert technical support. We acknowledge support from the QuantERA II Programme (project LEMAQUME) that has received funding from the European Union’s Horizon 2020 research and innovation programme under Grant Agreement No 101017733. Further, we would like to thank for support the UK funding agency EPSRC
under grants EP/W007444/1, EP/V035975/1, EP/V000624/1 and EP/X009491/1, the Leverhulme Trust (RPG-
2022-57), the EU Horizon 2020 FET-Open project TeQ (766900) and the EU Horizon Europe EIC Pathfinder project QuCoM (10032223), the DFG Project ID 390831469: EXC 2118 (PRISMA+ Cluster of Excellence), by the German Federal Ministry of Education and Research (BMBF) within the Quantumtechnologien program (Grant No. 13N15064), by the COST Action within the project COSMIC WISPers (Grant No. CA21106).

\paragraph*{Author Contributions:}
AV conceived the experiment together with HU and CT, and supervised the experiment. AV and FA designed and realized the experimental setup, with early contributions from HU and CT. FA performed the low temperature measurements, developed the data acquisition and data analysis tools and  performed data analysis. WJ and DB conceived the potential use of a levitated magnet as dark matter detector and estimated the sensitivity. All authors contributed to writing of the article. 

\paragraph*{Competing Interests:}
The authors declare no competing interests

\paragraph*{Data Availability:}
The experimental data used in the analysis will be available in future to any researcher for purposes of reproducing or extending the analysis.

\setcounter{equation}{0}
\setcounter{figure}{0}
\renewcommand{\theequation}{S\arabic{equation}}
\renewcommand{\thefigure}{S\arabic{figure}}

\newpage

\section*{SUPPLEMENTARY MATERIAL} 

\section*{Material and methods}

The superconducting trap is machined from a bulk rod of lead with $99.99\%$ purity. The trap bottom is hemispherical, in order to minimize the sensitivity of normal modes to trap tilts \cite{Vinante2020}. The nominal curvature radius is $2.5$ mm.
The micromagnet is an individual microsphere picked from a commercially available powder \cite{Magnequench}. This type of microspheres has been chosen for two reasons: it has been shown to maintain ferromagnetic properties at low temperature \cite{Vinante2020} and it is available as powder composed of nearly perfect microspheres with variable diameter.
The micromagnet is individually picked from the powder, magnetized in a $10$ T magnet to bring magnetization to the saturation limit and then placed on the bottom of the coil. It is then spontaneously levitated by Meissner effect when the setup is cooled below the critical temperature of lead ($T_c=7.1$ K). 
The pick-up coil and field coils are made each of 3 loops of NbTi wire would around a nylon support. All coils are 8-shaped and lie on a horizontal plane at an estimated distance of $2$ mm (pick-up coil) and $2.4$ mm (field coils) from the bottom of the trap. The outer diameter of the coils is $4$ mm.
The pick-up coil is connected to a commercial dc SQUID (Quantum Design) operated with its electronics (Quantum Design, model 550). The SQUID has input inductance $1.8$ $\mu$H and flux noise of $5$ $\mu\Phi_0/\sqrt{\mathrm{Hz}}$.
For the magnetic field resolution and ringdown measurements, the SQUID output signal is processed with a lock-in amplifier (Stanford Research, SR830).

\section*{Measurement of resonance frequencies and Q factor}

\begin{figure}[!t]
\centering
\includegraphics[width=0.5\linewidth]{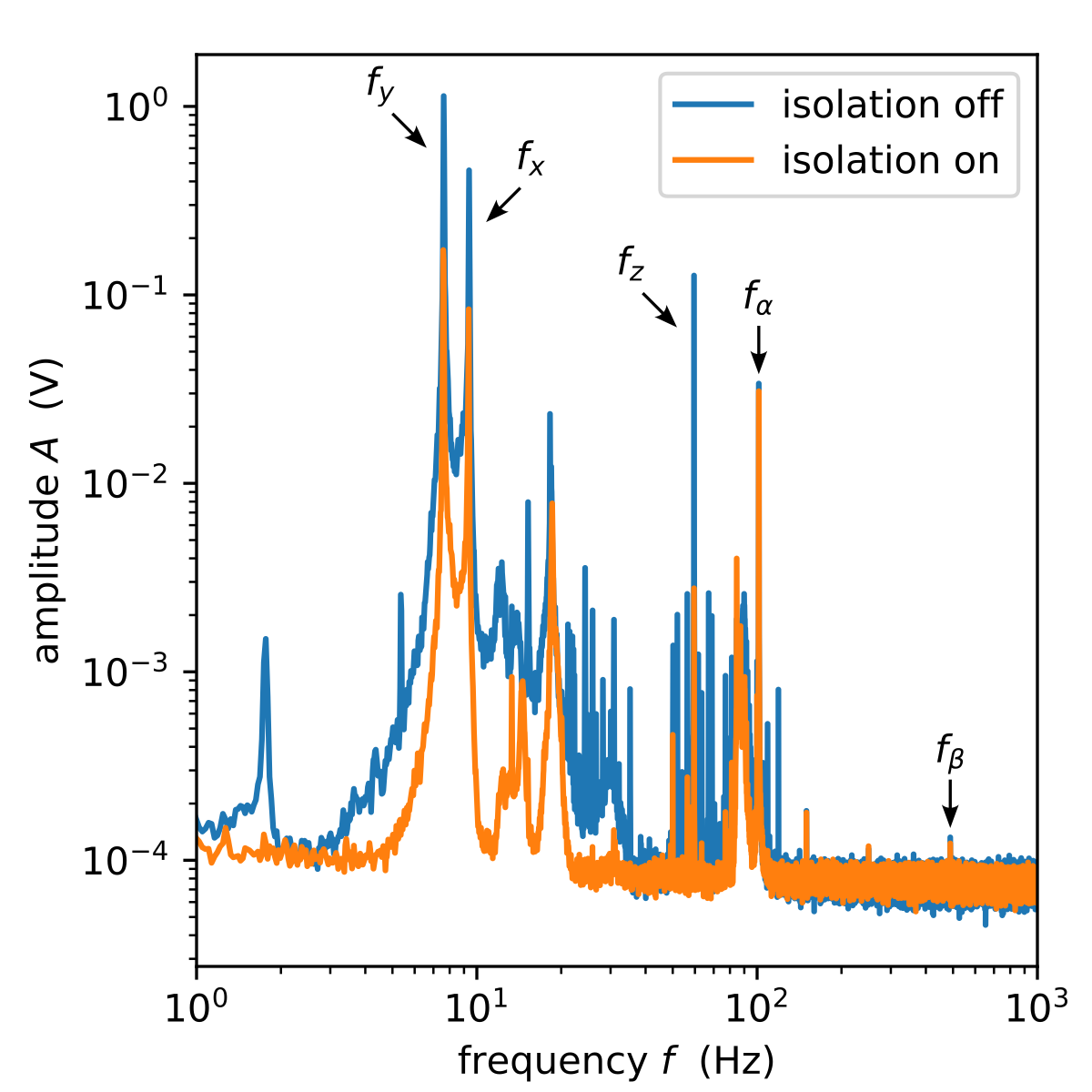}
\caption{Example of spectrum of the SQUID output signal. The five modes of the levitated magnet are identified by the high quality factors. Other spurious vibrational peaks are visible.}
\end{figure}

We detect five normal modes of the levitated microsphere, three translational ($x,y,z$) and two librational ($\alpha,\beta$).  As described in previous papers, the modes can be easily found in the power spectrum of the SQUID output signal, an example is shown in Fig. S1, due to their high quality factor $Q$, which scales with the inverse of the gas pressure $P$. Specific modes are identified by comparison with analytical and finite element modeling \cite{Vinante2020,Timberlake2019,Vinante2022}.

All measurements shown in this paper were acquired at a nominal pressure $P=1.1 \times 10^{-3}$ mbar of helium gas at the temperature $T=4.18$ K of the liquid helium cryostat. In order to accurately measure the resonance frequency $f_\alpha$ and the mechanical  factor $Q$ of a given mode we perform ringdown measurements. The specific mode is excited with the actuation coil and the exponential ringdown signal from each mode is acquired using a lock-in amplifier with reference frequency close to $f_\alpha$. The phase drift during the ringdown provides $f_\alpha$ with accuracy of $1$ mHz, while the $Q$ factor is evaluated as $Q=\pi f_\alpha \tau$ with $\tau$ ringdown time. $\tau$ is estimated from an exponential fit as shown in Fig. S2 with uncertainty better than $1\%$. For the mode $\alpha$ at the pressure $P=1.1 \times 10^{-3}$ mbar, corresponding to the magnetic field measurements discussed in this paper, we find $f_\alpha=137.625$ Hz and $Q=3.96 \times 10^4$.
\begin{figure}[!t]
\centering
\includegraphics[width=0.5\linewidth]{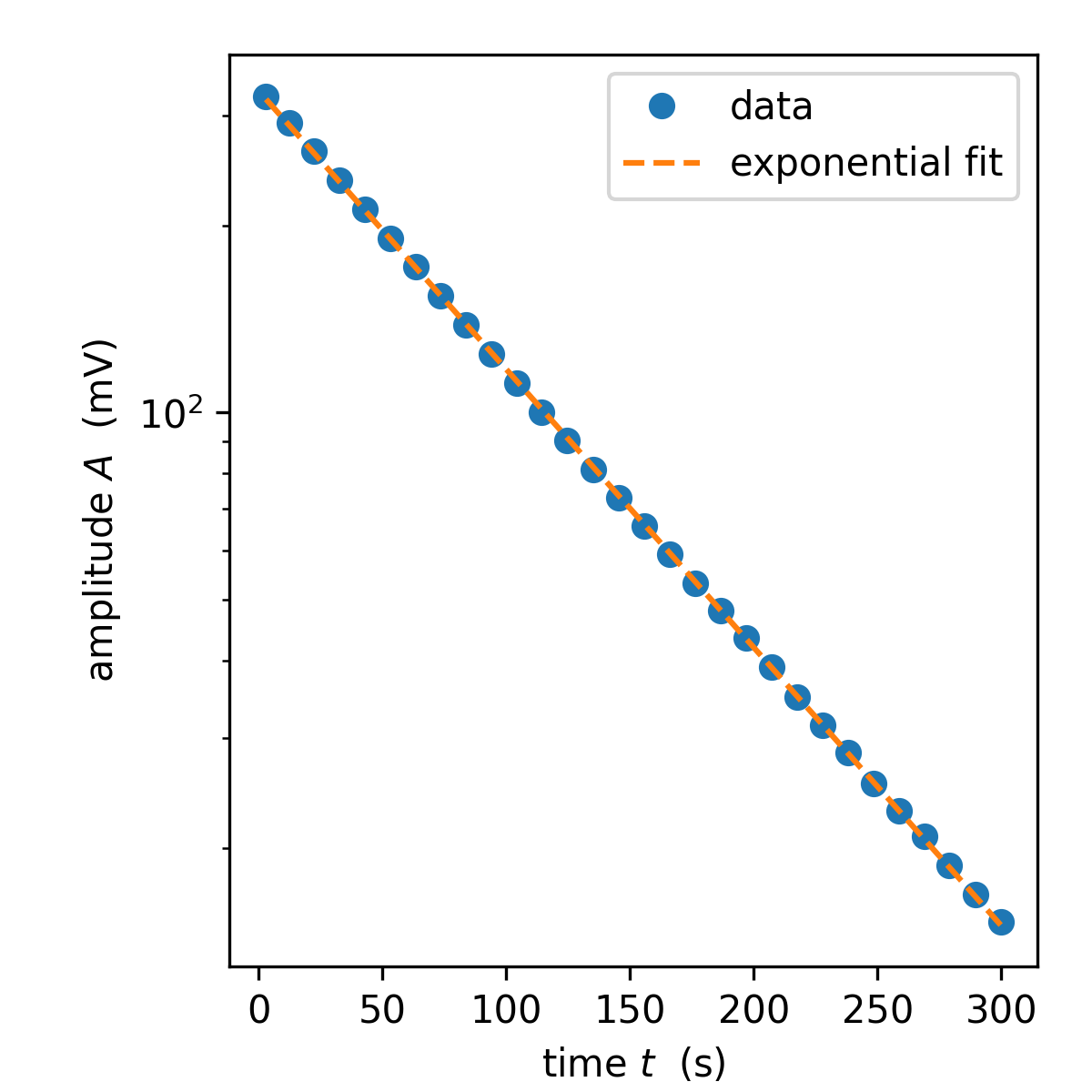}
\caption{Ringdown measurement at helium pressure $P=1.1\times10^{-3}$: amplitude of the $\alpha$ mode detected by the lock-in amplifier as function of time. The solid line is an exponential fit, providing the ringdown time $\tau$.}
\end{figure}

\section*{Determination of micromagnet parameters}

We determine the radius $R$ and the magnetization $M$ of the ferromagnetic microsphere in-situ from the resonance frequencies of the $z$ and $\beta$ modes. As discussed in refs. \cite{Vinante2020,Timberlake2019,Vinante2022} the problem of a magnet above an infinite superconducting plane can be modeled analytically using the image method. The potential energy is a function of the vertical coordinate $z$ and the azimuthal angle $\beta$, as well on the parameters of the sphere (radius $R$, density $\rho$, magnetization $M$) and on the gravity acceleration $g$.

Since our trap has a hemispherical bottom and the equilibrium position of the particle (estimated a posteriori as $z_0 \approx 270\,\mu$m), is much smaller than the radius of the trap hemisphere ($a=2.5$ mm), a spherical cavity inside a superconductor is a much better approximation than an infinite plane. The image method provides analytical solutions for this case as well \cite{Lin2006}. The potential energy can be written as:
\begin{equation}
U= \frac{\mu_0 \mu}{4 \pi} \frac{a^5}{\left(  a^2+r^2 \right) \left( a^2-r^2 \right)^3} \left( 1+\frac{a^2}{r^2} \mathrm{sin}^2 \beta \right) + m g z.
\end{equation}
Here, $r$ is the distance of the magnet center from the hemispherical trap center, $m=\rho V$ is the mass and $\mu=M V$ the magnetic dipole moment, where $V=4 \pi/3 R^3$ is the magnetic microsphere volume. The equilibrium height from the bottom $z_0=a-r$ and the equilibrium angle $\beta$ can be calculated numerically by minimization of $U$. As in the infinite plane case, the equilibrium orientation is on the horizontal plane, i.e., $\beta=0$.
Resonance frequencies of $z$ and $\beta$ modes can be calculated numerically from the second derivatives of $U$, the mass $m$ and the moment of inertia $I=2/5mR^2$:
\begin{align}
 f_z&=\frac{1}{2 \pi}\sqrt{\frac{1}{m}\frac{\partial^2 U}{\partial z^2}\bigg|_{z=z_0,\beta=0}} \\
  f_\beta&=\frac{1}{2 \pi}\sqrt{\frac{1}{I}\frac{\partial^2 U}{\partial \beta^2}\bigg|_{z=z_0,\beta=0}}
\end{align}

In our experiment we consider as unknown parameters the radius $R$ and the magnetization $M$. We consider known, within a systematic uncertainty, all other parameters. Gravity acceleration at our site in Trento (Italy) can be determined from gravitational reference data as $g=9.80674$ m/s$^2$ with negligible uncertainty. The density of the material is provided by the manufacturer at room temperature as $\rho=7430$ kg/m$^3$. We assume a conservative relative uncertainty of $5\%$ to take into account thermal contractions and possible oxidation of the surface. The trap radius is taken as $2.5$ mm as by design, with a relative uncertainty of $10 \% $ taking into account machining accuracy.

The resonance frequencies of $z$ and $\beta$ modes can be measured with very high precision, however we must take into account a major systematic uncertainty: Different cooldowns lead to different trapped field, which reflects into a different $\alpha$ mode frequency. This results also into a change of the $\beta$ mode. To understand this point, we observe that the trapped field component $B_0$ parallel to $\mu$ will determine a rotational stiffness $\kappa=\mu B_0$ which is the same in the two librational angles $\alpha$ and $\beta$. For the $\beta$ mode this implies a frequency shift to:
\begin{equation}
  f_{\beta}'=\sqrt{f_{\beta}^2+\frac{\kappa}{I}}=\sqrt{f_{\beta}^2+f_{\alpha}^2}.
\end{equation}
We have tested this relation by repeating several cooldowns and finding different pairs of frequencies $f_{\beta},f_{\alpha}$. Fig. S3 shows that the measured data are in agreement with Eq. (S4). A fit yields the intrinsic value $f_{\beta}=(478.3 \pm 0.5)$ Hz. A similar quadratic fit gives yields an intrinsic value for the $z$ mode frequency $f_z=(59.1 \pm 0.1)$ Hz. Note that these estimations, besides providing the intrinsic values of resonance frequencies at vanishing trapped field $B_0$, also give an experimental determination of the uncertainty based on reproducibility, that is much larger than the nominal precision of the frequency measurement from ringdown.
\begin{figure}[!t]
\centering
\includegraphics[width=0.5\linewidth]{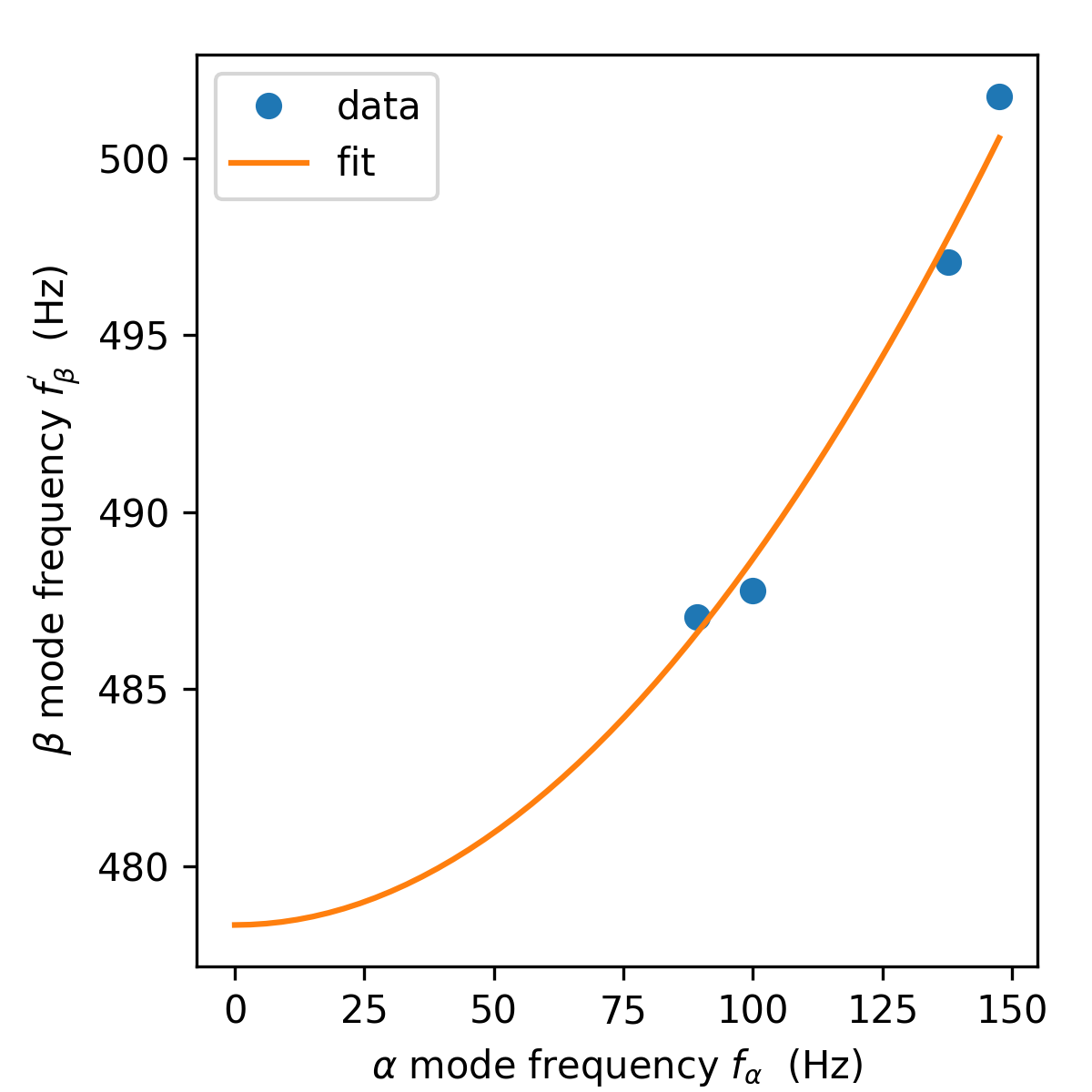}
\caption{Frequency of the $\beta$ mode as function of the frequency of the $\alpha$ mode for different cooldowns with the same micromagnet. A fit with Eq. (S4) provides our best estimation of the $\beta$ mode frequency at vanishing $B_0$ field.}
\end{figure}
Using all given values of $f_z, f_\beta, \rho,a,g $ with their respective uncertainty, we can invert numerically Eqs. (S2,S3) for $M$ and $R$ taking into account  error propagation. We find $R=(20.78\pm 0.20)$ $\mu$m and $M=(6.91 \pm 0.18) \times 10^5$ A/m, corresponding to a saturation field $B_\mathrm{s}=(0.87 \pm 0.02)$ T.
Using this value of $R$ we can estimate all other parameters, such as the volume $V$, the mass $m$ and the moment of inertia $I=2/5 m R^2$.

As a cross-check, we have performed a completely independent in-situ estimation of the radius $R$, based on a viscous damping measurement on the same micromagnet in a different cooldown. The measurement is performed at much higher gas pressure $P=1.6$ mbar, where we estimate a Knudsen number $K_n=l/R=0.05$, with $l$ the mean free path of helium gas atoms, implying that the motion of the microsphere is in the viscous regime. Under this condition, the viscous torque for rotational motion follows the Stokes law $T_v=-8 \pi \eta R^3 \Omega $, where $\Omega$ is the angular velocity. For a librational mode, this translates into a resonance linewidth:
\begin{equation}
  \Delta f =\frac{15 \eta}{2 \pi \rho R^2 }
\end{equation}
where $\eta$ is the gas viscosity.

\begin{figure}[!t]
\centering
\includegraphics[width=0.5\linewidth]{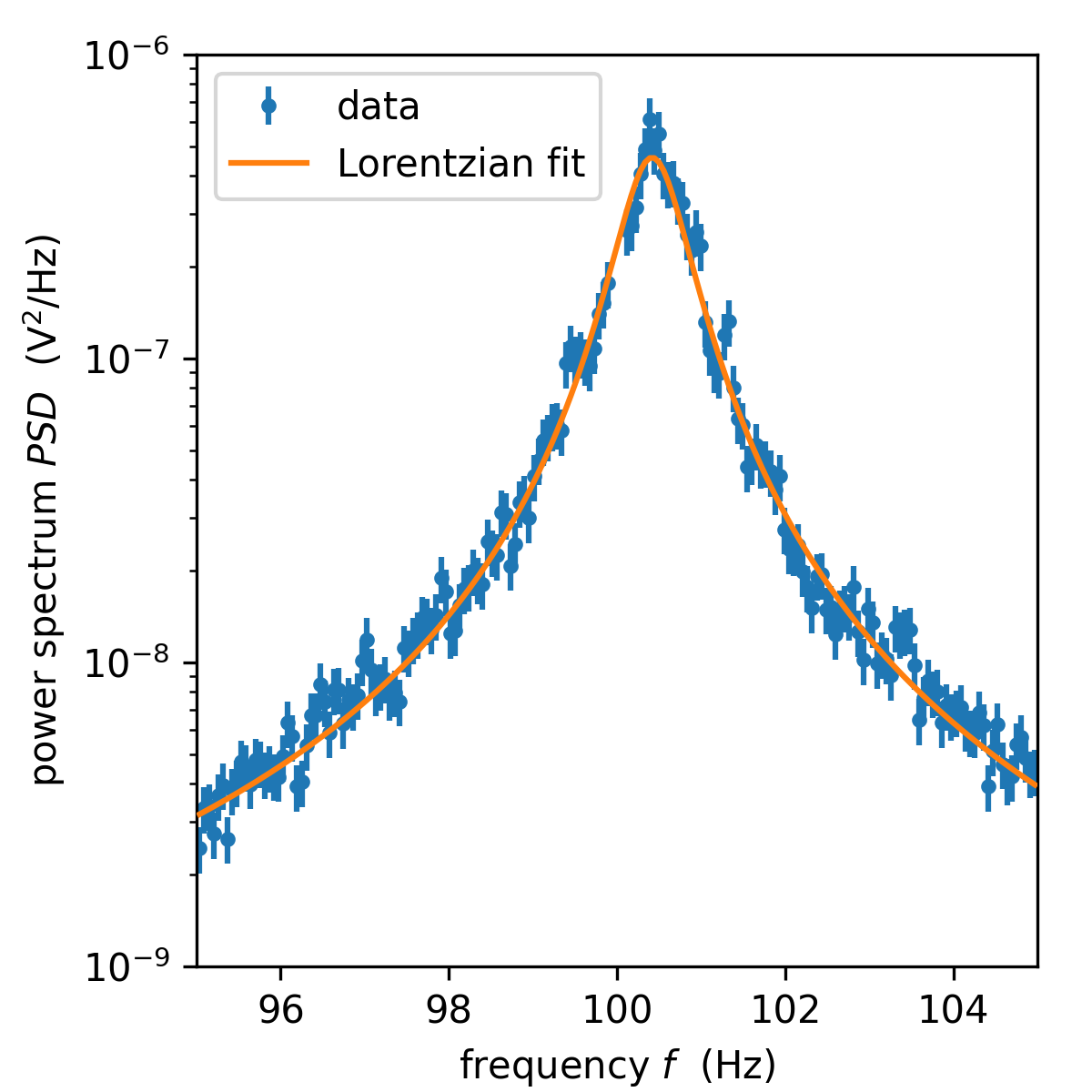}
\caption{Spectrum of $\alpha$ mode in a different cooldown at high pressure $P=1.6$ mbar in the viscous regime. The Lorentzian fit gives the linewidth $\Delta f = (0.852 \pm 0.026)$ Hz. Provided that the density of the microsphere and the gas viscosity are known, this measurement provides an estimation of the microsphere radius $R$.}
\end{figure}

In Fig. 4 we show a measurement of the linewidth in the power spectrum. From a Lorentzian fit we find $\Delta f=(0.852 \pm 0.026)$ Hz. Using for the viscosity the value $\eta=1.13 \times 10^{-6}$ Pa$\cdot$s found in literature for helium gas at $T=4.2$ K and pressure similar to the one present in our measurement \cite{Coremans1958}, and for the density the value $\rho=7430$ kg/m$^3$ with $5\%$ error as above, we obtain an independent estimation of the radius $R=(20.6 \pm 0.6)$ $\mu$m, compatible with the estimation of $R$ from the resonance frequencies. Note that this last estimation based on the viscosity has a larger error but is completely independent of magnetic properties and levitation modeling.

\section*{Data analysis and experimental uncertainties}
For the main measurements presented in Fig. 3, the data have been analyzed in the following way. The minimum detectable field $B_1$ for a given signal frequency $f$ is extracted from datasets as the one presented in Fig. 3A by setting the signal equal to the noise (SNR=1). This allows to define the magnetic field noise as $B_\mathrm{n}=B_{1,\mathrm{min}} = \lambda_1 I_{1,{\mathrm{SNR=1}}}$. 
Clearly, the minimum detectable field depends on the orientation with respect to the magnet, as the torque induced by $B_1$ is $\tau=\mu B_1 \mathrm{sin}\theta_1$. To estimate the magnetic field resolution for an optimally oriented field we multiply $B_{\mathrm{min}}$ by the factor $\mathrm{sin}\theta_1$. Finally, the minimum detectable field is converted into a PSD by taking into account that the noise power integrated over the measurement bandwidth $\Delta f$ of the lock-in amplifier, yielding $B_\mathrm{n}^2=S_B \Delta f$.

The uncertainty on the $S_B$ data points shown in Fig. 3B is found with standard error propagation of the uncertainties on the current $I_{1,{\mathrm{SNR=1}}}$, the factor $\lambda_1=B_1/I_1$ and the angle $\theta_1$.
The energy resolution quoted in the paper $E_\mathrm{R}=\left( 0.064 \pm 0.010 \right) \, \hbar$ is obtained from Eq. (1) by averaging the $S_B$ data in Fig. 3B around the minimum.

The orange line in Fig. 3B is the magnetic field noise PSD corresponding to the theoretical thermal noise. This is obtained by setting $S_B=S_\tau/\mu^2$ with the thermal torque noise $S_{\tau}=4 k_\mathrm{B} T I \omega_\alpha/Q$. The uncertainty is found from propagation of the uncertainties on $\mu$, $I$ and $Q$. The uncertainties on $T$ and $\omega_\alpha=2 \pi f_\alpha$ are negligible.

Finally the green line in Fig. 3B obtained from Eq. (1) by setting $E_R=1 \hbar$, and its uncertainty derives from the uncertainty on the volume $V$.

\section*{Axion dark matter limits}

In this section we summarize the procedure to derive the curves plotted in Fig. 4. 
The axion-electron coupling's Lagrangian is\cite{crescini2020axion} 
\begin{equation}
    L=\frac{g_{aee}}{2 m_e} \partial_\mu a(\bar{\psi}_e\gamma^\mu\gamma^5\psi_e)
\end{equation}
\noindent where $m_e$ is the mass of electron. In the non-relativistic limit, the axion gradient interaction with electron spin can be described as 
\begin{equation}
    H=\gamma_e \hbar \frac{g_{aee}}{2 e}\bm{\nabla} a \cdot \bm{\sigma}_e =\gamma_e\hbar \bm{\sigma}_e\cdot (\bm{B}_{eff})
\end{equation}
\noindent where $\bm{\sigma_e}$ is the spin operator of the electron, and $\bm{B}_{eff}=(g_{aee}/2e)\bm{\nabla}a$ is the Zeeman-like effective magnetic field.
First, we assume that axions constitute the totality of the galactic halo dark matter density, and assume the levitated magnet sensor's response to the magnetic field is primarily dominated by its spins, and neglect its orbital magnetic moments. In the non-relativistic limit, the axion dark matter is well described by a classical field 
\begin{equation}
    a(t)\approx a_0 \rm{cos}(\omega_{DM}t).
\end{equation}
Where $\omega_{DM}\approx m_{DM}c^2/\hbar$ is the axion Compton frequency and $a_0$ is axion field amplitude. $a_0$ relates to the dark matter density through 
\begin{equation}
    \rho_{DM}=\frac{1}{2}\frac{c^2}{\hbar^2}m^2_{DM} a_0^2
\end{equation}
From the above equations we can derrive $\nabla a$ in the form of Eq. (2) in the manuscript (in the main manuscript we use natural units and $\hbar=c=1$ is ignored there). 
The effective magnetic field felt by electron spins can then be calculated from Eq. (2) as $B_a={(2.1\times 10^{-8}\cdot g_{aee} \rm ) \, T}$. 
The coherence time of the axion field is taken as $t_\mathrm{coh}\simeq 10^6/{2\pi\omega_{DM}}$.

We then assume that at a given probe frequency our setup has a magnetic field noise limited by thermal noise, given by the relation $S_{B}=4 k_\mathrm{B} T I \omega_0/(\mu^2 Q)$
which further assumes optimal orientation.
In general for a full and more realistic modeling we should consider the readout noise from the SQUID, as well as other technical noise sources. However, for the parameters used here the thermal noise is always larger than the mechanical standard quantum limit, so that in principle we can always find a configuration where we are thermal noise limited. Therefore, we consider here for simplicity only thermal noise, leaving a complete and exhaustive noise modeling to a future technical paper.
We consider two configurations: a present one corresponds to the setup described in this paper, with $R=20$ $\mu$m, $T=4.2$ K, $Q=4 \times 10^4$, $f_0=\omega_0/2 \pi=100$ Hz and a future one with a larger magnet $R=2$ mm, $T=50$ mK, $Q=10^8$, $f_0=10$ Hz. The latter configuration is of course much more challenging, but the quoted parameters have been essentially achieved or approached in separate tests \cite{Vinante2020}.

The signal to noise ratio for detecting an axion signal at frequency $f$ is given, for short integration $t<t_\mathrm{coh}$, by $\mathrm{SNR}=(B_a^2/S_B)t$ see Ref.\cite{higgins2023maglev}. For longer integration time the averaging is incoherent and $\mathrm{SNR}=(B_a^2/S_B)\sqrt{t \, t_\mathrm{coh}}$. The red-dashed and green-dashed curves in Fig. 4 are derived by assuming an integration time of 1 year at SNR=1, For the frequency below 0.01\,Hz the coherence time is longer than integration time and the sensitivity is flat; while for the frequency above 0.1\,Hz, the coherence time is shorter than one year and the sensitivity decreases with increasing frequency.

\end{document}